\documentclass[journal=jpcbfk,manuscript=article,layout=traditional]{achemso}

\usepackage{amsmath}

\def\i{{\mathrm i}}

\newcommand{\bfr}{\textbf{r}}
\newcommand{\onehalf}{\frac{1}{2}}

\SectionNumbersOn

\author{D.\ Hofmann-Mees$^1$, H. Appel$^{2,3,4}$, M. Di Ventra$^3$, and S. K\"ummel$^1$}
\affiliation{$^1$ Theoretical Physics IV, University of Bayreuth, D-95440 Bayreuth, Germany}
\affiliation{$^2$ Fritz-Haber-Institut der Max-Planck-Gesellschaft, Faradayweg 4-6, D-14195 Berlin, Germany}
\affiliation{$^3$ Department of Physics, University of California San Diego, La Jolla, CA 92093, USA}
\affiliation{$^4$ European Theoretical Spectroscopy Facility}
\email{stephan.kuemmel@uni-bayreuth.de;  Phone: +49 921 553220}

\title{Determining Excitation-Energy Transfer Times and Mechanisms from Stochastic Time-Dependent Density Functional Theory}

\begin{document}

\begin{abstract}

We developed an approach for calculating excitation-energy 
transfer times in supermolecular arrangements based on 
stochastic time-dependent density functional theory (STDDFT). 
The combination of real-time propagation and the stochastic Schr\"odinger 
equation with a Kohn-Sham Hamiltonian allows for simulating how an excitation 
spreads through an assembly of molecular systems. The influence that approximations, 
such as the dipole-dipole coupling approximation of F\"orster theory, have on 
energy-transfer times can be checked explicitly. As a first application of our 
approach we investigate a light-harvesting inspired model ring system, calculating 
the time it takes for an excitation to travel from one side of the ring to the 
opposite side under ideal and perturbed conditions. Among other things we find 
that completely removing a molecule from the ring may inhibit energy transfer 
less than having an energetically detuned molecule in the ring. In addition, 
F\"orster's dipole coupling approximation may overestimate excitation-energy 
transfer efficiency noticeably.

\end{abstract}


\section{Introduction}

Electronic excitation-energy transfer (EET) after light absorption is one of the key processes in the natural light-harvesting (LH) event and a prerequisite for charge generation in the LH reaction center \cite{kuhlbrandt:497,cheng:241}. Hence, it enables efficient energy recovery in biological systems. The efficiency of the light-harvesting process is determined by the rates of charge transfer and EET of many single transfer steps that all contribute to the overall mechanism. In multichromophoric supermolecules such as LH systems, the rates are affected by a number of different properties and phenomena: the electronic structure of the single chromophores, the electronic coupling between different system components, the geometry and arrangement of all constituents, the energetic and position (dis)order, and the interplay with the environment.

Recent years have seen an intense debate about the importance of intrinsic quantum mechanical effects for the  LH process \cite{ritz:243, fleming:256, rebentrost:9942, olaya-castro:49}, specifically the existence of long-lived quantum coherences \cite{engel:782, fleming:27, struempfer:536}. 
The character of these coherences, e.g., their electronic or vibronic nature, is still controversial \cite{turner:1904, yuen-zhou:234501}.
Quantum-coherence effects have also been found in artificial $\pi$-conjugated polymers \cite{bredas:48, collini:369}. It has been argued that quantum-mechanical interference between different energy-transfer pathways may improve the efficiency of intrachain energy migration \cite{bredas:48}.
Thus, understanding the design of natural light-harvesting complexes may guide the future design of artificial organic devices \cite{scholes:763, fleming:27,harel}. 

Coherent energy transfer and the role of a system's environment can theoretically be studied using the density-matrix formalism. The models of Haken and Stobl, the model of Redfield, polaron modifications, or related theories have been employed \cite{jang:319, yang:355, jang:101104, olsina:1765, ishizaki:234110, ishizaki:234111, kolli:154112, rebentrost:9942}. These concepts typically rely on input from electronic-structure theory or experiment. In the present work we present an alternative approach that addresses the energy-transfer problem directly at the level of electronic-structure theory by employing stochastic time-dependent density functional theory (STDDFT)~\cite{ventra:226403, agosta:165105, appel:212303, appel:27, biele:2012, appel:2012, bushong:395214, pershin:054302, agosta:2009} which is an extension of standard time-dependent density functional theory (TDDFT) in real time \cite{runge:997, gross:81, yabana:4484, yabana:55, marques:427, marques:2012}, specifically formulated for open
quantum systems.

Using TDDFT appears promising because it allows for calculating the electronic structure, the excitations, and also the couplings between the constituents of a supermolecular structure in principle exactly. TDDFT in practice can address multichromophoric systems of experimental relevance at a first-principles level at reasonable computational costs. The price one has to pay are the limitations of the available approximations for the exchange-correlation potential on which the predictive power of TDDFT calculations strongly depends. Luckily, in many energy-transfer situations long-range exchange and correlation (xc) do not play a decisive role as the intermolecular separations are large \cite{neugebauer2007,hofmann:012509}. Furthermore, recent progress in developing density functionals that can even describe charge-transfer situations reliably also allows for accurately describing excitations of complicated nature in complex systems \cite{yanai:51, chai:084106, peach2008, stein:2818,KAK09,karolewski:151101,
hofmann:146401,tdsicjcp,tuningperspective,sicinteger}. TDDFT with modern functionals thus appears as an ideal tool for studying energy-transfer processes on a first-principles basis.

However, there is still a conceptual hurdle that must be overcome. In most situations of interest the electronically active parts of a system interact with an environment, e.g., the phonons of a solvent or matrix. Taking all these degrees of freedom explicitly into account in a TDDFT calculation is presently computationally impossible, and will probably be so for years to come. Therefore, an effective description that traces out the environment degrees of freedom is highly desirable. However, standard TDDFT is formulated using the closed quantum-system time-dependent Kohn-Sham (TDKS) equations that evolve coherently in time. In order to investigate the real-time dynamics of a system in contact with a dissipative bath one needs to go beyond standard TDDFT. Using open quantum-system schemes \cite{burke:146803, ventra:226403, agosta:165105, appel:212303, appel:27, biele:2012, appel:2012, bushong:395214, pershin:054302, yuen-zhou:4509, agosta:2009, yuen-zhou:043001, tempel:130, tempel:074116, zheng:14358, tempel:2012} is 
an attractive option. The real-time TDDFT scheme for investigating EET processes that we present here employs the stochastic TDKS equations \cite{ventra:226403, agosta:165105, appel:212303, appel:27, biele:2012, appel:2012}. We introduce a 
pragmatically motivated effective bath operator that couples to the dipole moment of the supermolecule or appropriate subsystems. 
As a first demonstration of our scheme we investigate a ring configuration of model molecules that is schematically inspired by nature's circular LH complexes. We examine how different ways of treating the intermolecular coupling, e.g., using F\"orster's dipole coupling approximation or not, and how energetic disorder influence energy-transfer times.
In this first study we use the bath operator for determining the time that an excitation takes to spread in the system. 

The paper is organized as follows. In \ref{theory_section} and the corresponding \ref{sp_bath_op_appendix} we outline the stochastic TDKS approach and introduce a bath operator that couples to the dipole moment. Our ring-like model system is introduced in  \ref{model_system_section}. In this section we also test the bath operator and explain the idea of using it for determining the effective time that an excitation travels in a supermolecular system. In  \ref{results_section} we report how variations in the electronic structure and how different intermolecular coupling mechanisms influence EET times. A summary and conclusions are offered in  \ref{summary_section}. Finally, \ref{model_system_appendix} and \ref{coupling_sec} provide technical details about our electronic structure setup and our method for extracting coupling strengths from real-time TDDFT, respectively.

\section{Dissipation in real-time time-dependent density functional theory}
\label{theory_section}

\subsection{Stochastic Schr\"odinger equation}

It is one of the basic ideas of open quantum-system approaches to split the entire Hamiltonian into system, bath, and system-bath interaction contributions, i.e.,
\begin{equation}
 \hat{H} = \hat{H}_\mathrm{S} \otimes \hat{I}_\mathrm{B} + \hat{I}_\mathrm{S} \otimes \hat{H}_\mathrm{B} + \lambda \hat{H}_\mathrm{SB}.
\end{equation}
$\hat{I}_\mathrm{S}$ and $\hat{I}_\mathrm{B}$ denote identity operators in the system (S) and bath (B) Hilbert spaces. The system includes all dynamics and observables of, e.g., one molecular complex. The bath and system-bath coupling describe the environment, e.g., other surrounding molecules, and their interactions with the system. The system degrees of freedom can thus be coupled to a bosonic environment. The system itself is described by the electronic many-particle Hamiltonian
\begin{equation}
\begin{split}
  \hat{H}_\mathrm{S} = & \sum_i \left[ \frac{[\hat{\mathbf{p}}_i+e\hat{\mathbf{A}}_\mathrm{ext}(\hat{\bfr}_i,t)]^2}{2m} + \hat{v}_\mathrm{ext}(\hat{\bfr}_i,t) \right] \\
  &  + \sum_{i<j} \hat{W}(\hat{\bfr}_i-\hat{\bfr}_j),
\end{split}
\end{equation}
where $\mathbf{A}_\mathrm{ext}$ is an external vector potential, $v_\mathrm{ext}$ a scalar external potential, and $\hat{W}$ describes the particle-particle interaction. The environment, represented by $\hat{H}_\mathrm{B}$, induces a fluctuating force that drives the system via the system-bath interaction
\begin{equation}
 \hat{H}_\mathrm{SB} = \sum_\alpha \hat{S}_\alpha \otimes \hat{B}_\alpha.
\end{equation}
This interaction is expressed in terms of many-particle operators $\hat{S}_\alpha$ and $\hat{B}_\alpha$, where $\hat{S}_\alpha$ operates on the system degrees, $\hat{B}_\alpha$ on the bath degrees of freedom, and $\alpha$ denotes different possible system-bath coupling mechanisms. The bath may have a complex structure and not all of its microscopic details are relevant for the system dynamics. The driving force that is induced by the bath may, therefore, be subsumed by stochastic fluctuation and dissipation contributions that can usually be characterized by mean values and correlation functions \cite{gaspard:5676}. The parameter $\lambda$ determines the strength of the system-bath interaction and can be used as an expansion parameter in the weak-coupling regime.

The dynamics of a quantum system in contact with an external bath can be described \cite{dalibard:580, gaspard:5676, breuer:2006, kampen:2007, weiss:2008, biele:2012, appel:2012} either by quantum master equations or by the stochastic Schr\"odinger equation (SSE). The latter uses a statistical ensemble of state vectors to represent the open quantum-system dynamics directly on the level of wave functions.

Throughout this work we use the SSE in the Born-Markov limit \cite{dalibard:580, gaspard:5676, breuer:2006, kampen:2007, weiss:2008, biele:2012, appel:2012}
\begin{equation}
\begin{split}
 \label{stSEq}
 \mathrm i \partial_t \Psi_\mathrm{S}(t) = & \hat{H}_\mathrm{S}(t)\Psi_\mathrm{S}(t)-  \frac{\mathrm i}{2}\sum_\alpha \hat{S}_\alpha^\dagger \hat{S}_\alpha \Psi_\mathrm{S}(t) \\
    &  + \sum_\alpha l_\alpha (t) \hat{S}_\alpha \Psi_\mathrm{S}(t),
\end{split}
\end{equation}
where $l_\alpha (t)$ are stochastic processes with vanishing ensemble average and $\delta$-time-correlation
\begin{equation}
 \overline{l_\alpha (t)} = 0,\ \overline{l_\alpha (t) l_\beta (t')} = 0,\ \overline{l_\alpha^* (t) l_\beta (t')} = \delta_{\alpha\beta}\delta(t-t').
\end{equation}
The bar denotes the statistical average over an ensemble of stochastic processes. For the sake of convenience the coupling strength parameter $\lambda$ here has been absorbed into the bath operator $\hat{S}_\alpha$. The first term of the SSE\ (\ref{stSEq}) determines the usual unitary system evolution under the action of the system Hamiltonian $\hat{H}_\mathrm{S}$. Although the SSE employs the system Hilbert space only, the coupling to the bath is still included by the second term that describes dissipation effects due to the system-bath interaction. Finally, the third term introduces fluctuations in the time evolution: Although the dissipative term causes the probability density to decay in time, the norm of the state vector $\Psi(t)$ averaged over a statistical ensemble of realizations is conserved up to fourth order in the system-bath coupling parameter $\lambda$. 
In the following, we focus on a single bath operator $\hat{S}$ and can thus omit the index of the bath operator.

As a result of the stochastic nature of  (\ref{stSEq}) the system wave function can not be simulated by a single evolution of the SSE but needs to be represented by a statistical ensemble of wave functions $\{\Psi_s(t)\}$.
Accordingly, starting from a pure initial state, the full time evolution of expectation values
\begin{equation}
 \overline{O_\mathrm{S}}(t) = \overline{\langle\hat{O}_\mathrm{S}\rangle} = \overline{\langle\Psi(t)|\hat{O}_\mathrm{S}|\Psi(t)\rangle}
 \label{observables_st}
\end{equation}
of physical observables needs to be calculated from the statistical average over all ensemble members as indicated by the bar. A reliable computation of smooth observables requires a large enough set of stochastic realizations.

Working with the SSE tremendously reduces the complexity of the problem one has to solve. However, it introduces the question of choosing appropriate bath operators and it does not answer the question of how to describe the system itself ($\hat{H}_\mathrm{S}$). In the following we first address the latter question and then the former.

\subsection{Stochastic Schr\"odinger equation and Kohn-Sham density functional theory}

It has been shown that the open quantum-system SSE approach can be combined with TD current density functional theory (TDCDFT) \cite{ventra:226403, agosta:165105}, and the applicability of the approach has been discussed in Refs.\ \cite{appel:27, appel:2012, biele:2012}. In contrast to standard TD(C)DFT, the open quantum-system approach operates on ensemble-averaged quantities. It requires the ensemble-averaged particle density
\begin{equation}
 \overline{n(\bfr,t)} = \overline{\langle \hat{n}(\bfr) \rangle},
\end{equation}
where the density operator is defined as
\begin{equation}
 \hat{n}(\bfr) = \sum_i \delta(\bfr - \hat{\bfr}_i),
\end{equation}
and the ensemble-averaged current density
\begin{equation}
 \overline{\mathbf{j}(\bfr,t)} = \overline{\langle \hat{\mathbf{j}}(\bfr,t) \rangle},
\end{equation}
where the current operator reads
\begin{equation}
 \hat{\mathbf{j}}(\bfr,t) = \onehalf \sum_i \left\{\delta(\bfr - \hat{\bfr}_i), \frac{\hat{\mathbf{p}}_i+e\hat{\mathbf{A}}_\mathrm{ext}(\hat{\bfr}_i,t)}{m} \right\}
\end{equation}
and $\{.\,,.\}$ denotes the anticommutator bracket. 
The KS Hamiltonian reads
\begin{equation}
\begin{split}
 \hat{H}_\mathrm{KS} = \sum_i & \left[ \frac{[\hat{\mathbf{p}}_i+e\hat{\mathbf{A}}_\mathrm{ext}(\hat{\bfr}_i,t) +e\hat{\mathbf{A}}_\mathrm{xc}(\hat{\bfr}_i,t)]^2}{2} \right. \\
 & \left. + \hat{v}_\mathrm{ext}(\hat{\bfr}_i,t) + \hat{v}_\mathrm{H}(\hat{\bfr}_i,t) \right],
\end{split}
\end{equation}
with the external vector potential $\mathbf{A}_\mathrm{ext}(\bfr,t)$,
the xc vector potential $\mathbf{A}_\mathrm{xc}(\bfr,t)$, and the Hartree potential $v_\mathrm{H}(\bfr,t)$. The xc vector potential in the open quantum-system approach may in general depend on $\overline{\mathbf{j}(\bfr,t)}$, the initial state $\Phi_\mathrm{KS}(\bfr,t=0)$, and the bath operator $\hat{S}$. The system dynamics can be calculated from the KS Slater determinant $\Phi_\mathrm{KS}(\bfr,t)$ evolving according to the open system KS equation \cite{ventra:226403}
\begin{equation}
 \mathrm i \partial_t \Phi_\mathrm{KS}(t) = \hat{H}_\mathrm{KS}\Phi_\mathrm{KS}(t)-\frac{\mathrm i}{2} \hat{S}^\dagger \hat{S} \Phi_\mathrm{KS}(t) + l (t) \hat{S} \Phi_\mathrm{KS}(t).
 \label{stKSEq}
\end{equation}
For practical calculations one needs to rely on existing approximations for the xc vector potential as the true $\mathbf{A}_\mathrm{xc}(\bfr,t)$, especially in open quantum systems, is not known \cite{ventra:226403}.

Practical applications of stochastic TDDFT so far rely on approximate realizations of the stochastic formalism in KS TDDFT instead of TDCDFT. In Refs.\ \cite{appel:212303, appel:27} the system Hamiltonian is approximated by a standard TDDFT Hamiltonian
\begin{equation}
 \hat{H}_\mathrm{KS} = \sum_i \underbrace{\left[ -\frac{\hat{\nabla}_i^2}{2}
  + \hat{v}_\mathrm{ext}(\hat{\bfr}_i,t) + \hat{v}_\mathrm{H}(\hat{\bfr}_i,t) + \hat{v}_\mathrm{xc}(\hat{\bfr}_i,t)\right]}_{=:\hat{h}_\mathrm{KS}(\hat{\bfr}_i,t)}
\end{equation}
and used in the open quantum-system KS equation (\ref{stKSEq}) together with a heuristic bath operator \cite{pershin:054302}. Here, we apply an open quantum-system KS scheme of the same 
pragmatic kind. Exchange and correlation are described by the adiabatic local density approximation. A detailed discussion of the difficulties of establishing a DFT Hamiltonian in the open quantum-system framework, addressing, e.g., issues such as the $v$-representability problem of the current density \cite{agosta:245103}, is given in \cite{hofmann:2012}.

Going over to standard TDDFT and letting bath operators act on the TD KS states implies possibly far reaching approximations, because neither is the KS wavefunction necessarily a good model for the true wavefunction, nor is it clear that xc effects in the system-bath couplings can be neglected. With respect to the first issue there is some hope that at least for many organic systems that are of interest in the light-harvesting context the approximation is not too bad, because the wavefunction of many organic semiconductors is dominated by a single Slater determinant \cite{ksvsgks}. This is reflected, e.g., in angular resolved photoemission experiments 
that reveal very clear signatures of single molecular orbitals \cite{puschnig,dauth}. Also for the model dimers that we study in this paper correlation is weak and one can hope that it is not too unrealistic to rely on the KS states. Evaluating the second issue, i.e., the importance of a more complete description of the system bath xc coupling, goes beyond the aims of the 
present manuscript in which the focus lies on exploring a transparent, pragmatically motivated bath operator that is described in the next section.

\subsection{The bath operator}
\label{bath_op_sec}

The choice of the bath operator is the second ingredient needed for defining the open quantum-system scheme. Our aim is a bath operator that introduces dissipation by taking the excited quantum system back to its ground state. Thus, we define the bath operator as a projector of the form
\begin{equation}
 \hat{S} = \sqrt{\gamma} M[n(\bfr,t)] |\Phi_\mathrm{KS}(t_0)\rangle \langle\Phi_\mathrm{KS}(t)|.
 \label{bath_op_scaled}
\end{equation}
It includes two scaling factors $\sqrt{\gamma}$ and $M[n(\bfr,t)]$. Choosing $M=1$ would amount to a de-excitation of the entire system with a decay rate $\gamma$ and a decay time constant $\tau=1/\gamma$. 

The aim of our present study is to explore the possibilities that the open quantum-system KS equations offer in the field of molecular excitation-energy transfer. With a density-dependent factor $M[n(\bfr,t)]$, the bath operator (\ref{bath_op_scaled}) allows for modeling a broad range of physical situations that can be of relevance in this context. Here we specifically want to investigate a dissipation mechanism that depends on the (transition) dipole moment of the excited system, i.e., we want to choose $M$ proportional to the time variation of the dipole moment $\mathbf{d}(t) = \int \bfr n(\bfr,t)\,d^3 r$:  $M[n(\bfr,t)] \propto |\mathbf{d}(t) - \mathbf{d}(t_0)|$. This form of the bath operator is not derived ab initio, but is motivated from physical considerations. We are here thinking about describing situations where several subsystems (e.g.\ several chromophores) can exchange energy with each other and with a surrounding environment (e.g.\ a matrix embedding the chromophores). When a subsystem is 
strongly excited (indicated by a large induced dipole moment) it should transfer more energy to the environment than a weakly excited subsystem. 
The operator $M$ deliberately models only the system part of the system-bath interaction as the bath side of the coupling mechanism is subsumed in the effective decay rate.
An accurate description of chromophore-matrix interactions in future work may of course need to employ more sophisticated bath operators, but in this first exploratory study here we want to test whether the transparent, dipole-dependent mechanism can be realized.

\subsection{Simulation algorithm}
\label{simu_algo}

Having introduced the stochastic TDDFT formalism and a model bath operator, we finally turn our attention to the practical algorithm that we use to solve the open system KS equations. Our simulations are conducted with the quantum-jump algorithm \cite{dalibard:580, gardiner:4363, breuer:428, breuer:2006} that has been introduced in the context of open quantum-system KS equations in Ref.\ \cite{appel:27}. The algorithm is implemented in the Bayreuth version \cite{mundt:035413} of the PARSEC
real-space 
program package \cite{parsec} and rests upon the single-particle bath operator
\begin{equation}
 \hat{s}_i = \sqrt{\gamma} M[n(\bfr,t)] | \varphi_i(t_0)\rangle \langle \varphi_i(t) |.
 \label{KS_bath_op}
\end{equation}
We demonstrate in \ref{sp_bath_op_appendix} that using this bath operator in the following set of quantum-jump single-particle equations is equivalent to using $\hat{S}$ in the corresponding algorithm for solving  (\ref{stKSEq}).
The quantum-jump algorithm relies on a piecewise deterministic evolution \cite{appel:27,breuer:2006} of the set of norm-preserving Schr\"odinger equations
\begin{equation}
 \mathrm i \partial_t \varphi_i(t) = \hat{h}_\mathrm{KS}\varphi_i(t)-\frac{\mathrm i}{2} \hat{s}^\dagger_i \hat{s}_i \varphi_i(t) + \frac{\mathrm i}{2} |\hat{s}_i\varphi_i(t)|^2 \varphi_i(t),
 \label{stSEq_norm_cons}
\end{equation}
that are interrupted by quantum jumps. The quantum jumps represent the non-deterministic action of the bath operator.
Occurrence of a quantum jump at a given time means that at this time the bath operator transfers the system to a different state according to its projector part. Here, the TD orbitals are transferred to the corresponding ground-state orbitals.
The points of time where such a jump occurs are determined by a random process according to a waiting-time distribution. However, as this waiting-time distribution is not known beforehand, it needs to be determined alongside the actual propagation of the open quantum KS system (for details, see Ref.\ \cite{appel:27}): The waiting-time distribution is determined from the decay of the norm
\begin{equation}
 \eta(t)=\frac{1}{N}\sum_i \int |\varphi_i^\mathrm{aux}(\bfr,t)|^2\,d^3 r
\end{equation}
of an auxiliary system of $N$ particles evolving according to the non-norm conserving equation
\begin{equation}
 \mathrm i \partial_t \varphi_i^\mathrm{aux}(t) = \hat{h}_\mathrm{KS}\varphi_i^\mathrm{aux}(t)-\frac{\mathrm i}{2} \hat{s}^\dagger_i \hat{s}_i\varphi_i^\mathrm{aux}(t).
 \label{stSEq_norm_decay}
\end{equation}
One obtains a single waiting time by drawing a random number in the interval [0,1] and choosing as the quantum-jump time the time $T$ at which $\eta(T)$ drops below this number. The waiting-time distribution is then calculated from many of such single jump times.

In this algorithm our choice of bath operator leads to an additional beneficial property. When no additional external perturbations act after the initial excitation, the bath mechanism relaxes the system to its ground state, and the system remains in the ground state after each quantum jump. Therefore, the time evolution of all members of the stochastic ensemble follows the same pattern: The KS system first evolves deterministically until a quantum jump occurs and then -- after the system has jumped back to the ground state -- the KS orbitals propagate trivially with TD phases following from the ground-state KS eigenvalues. Therefore, the full statistical ensemble can be generated from a single deterministic evolution of  (\ref{stSEq_norm_cons}) alongside with the determination of the waiting-time distribution. The complete simulation therefore proceeds as follows: (i) Calculate a long enough deterministic evolution together with the norm decay of the auxiliary system, i.e., solve Eqs. (\ref{stSEq_norm_cons}) and 
(\ref{stSEq_norm_decay}). (ii) Draw a large number of random numbers, find the corresponding quantum-jump times, and determine the waiting-time distribution. (iii) For each quantum-jump
time construct the corresponding trajectory by combining the time evolution as calculated in step (i) (used for the times smaller than the jump time) with ground-state values (used for the times larger than the jump time). (iv) Calculate the physical quantities of interest by averaging (cf.\  (\ref{observables_st})) the observables over the ensemble of trajectories that has been obtained in step (iii).

As an aside we note that an explicit orthogonalization of the orbitals as done in Ref.\ \cite{appel:27} is not needed here because the bath operator preserves the orthogonality of the KS orbitals.

\section{Model system and bath coupling mechanism}
\label{model_system_section}

\subsection{Model setup}

\begin{figure}
  \includegraphics[width=8.5cm]{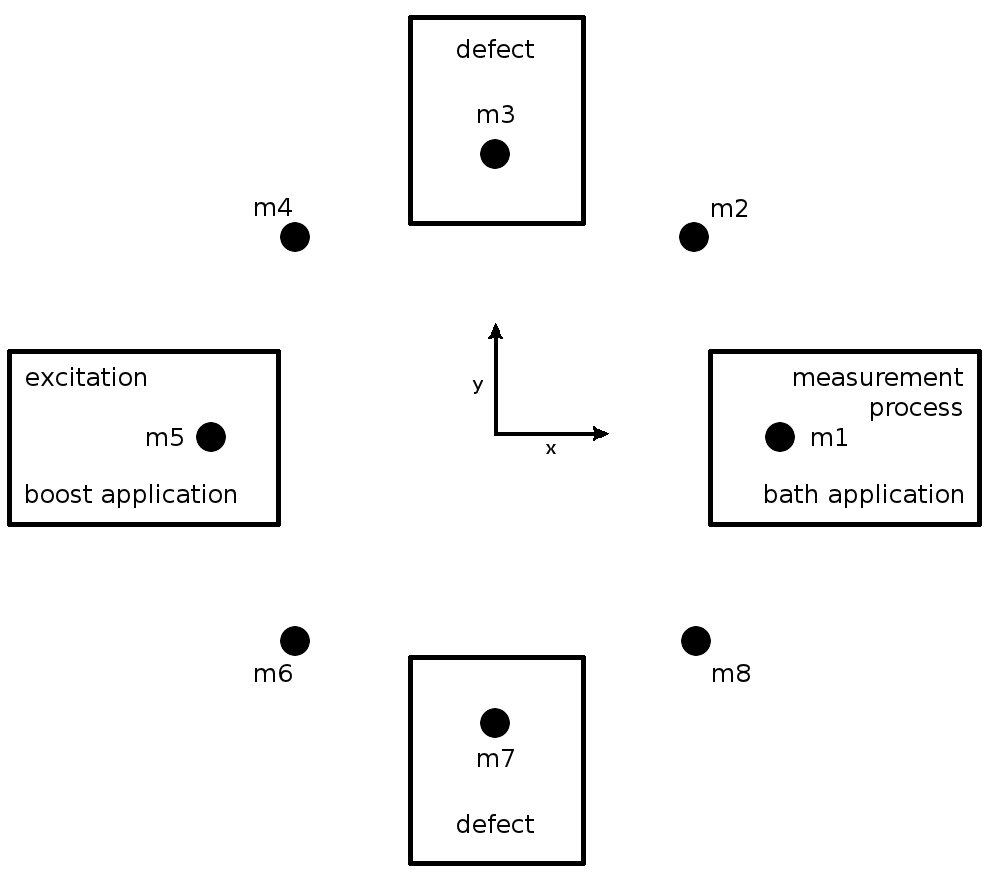}
  \caption{Schematic illustration of the ring-like model system consisting of eight individual molecules m1, m2, \ldots, m8. We determine how excitation energy spreads from m5 to m1 as described in the main text. For investigating the effects of changes in the intermolecular coupling we introduce defects in the molecules m3 and m7. See \ref{model_system_appendix} for further details.}
\label{supermol_arrangement}
\end{figure}

In the following we use the approach that we introduced in the previous sections for investigating energy transfer in a model system of circularly arranged molecules. The geometric arrangement of the molecules is designed to schematically resemble circular LH complexes of the antenna system of LH organisms. Our aim is to study the influence of electronic-structure properties on the EET time scales and pathways in such supermolecules. In this first study we focus on the influence of the coupling between the molecules that build the supermolecule.

The setup consists of eight equidistantly spaced molecules, see  \ref{supermol_arrangement}. Starting from the ground state we excite one of the molecules, in the following labeled m5. 
This excitation of the one molecule m5 is numerically realized by applying the boost excitation that is described below only to the (45-degree) sector of the grid that is associated with m5. In other words, all eight molecules are placed on the same large threedimensional grid during the simulation, but for excitation and bath action we associate each molecule with a certain part of the grid, namely the 45-degree sector in which it lies.

We simulate the excitation by a momentum boost of the type that is frequently used in real-time TDDFT \cite{yabana:4484,yabana:55,applphysb,octopus,adiaexPCCP}.
The boost has dipole character, and this conforms with our aim of modeling optical dipole excitations. However, whereas the external potential for a laser excitation (assuming a classical, non-quantized electromagnetic field) would correspond to a TD oscillating dipole potential with a TD envelope function, the boost is instantaneous. It can be understood as a ``spectrally absolutely 
broad'' excitation that can couple to all eigenfrequencies of a system and is therefore not a model for a laser field. However, for the purposes of our present study we deliberately choose this type of excitation because it allows for a precise definition of the time at which the excitation begins and a corresponding definition of an energy-transfer time. Furthermore, in the model system that we study here the spectrum of an individual molecule for a given excitation-polarization direction is strongly dominated by a single excitation. Therefore, the boost can be interpreted as instantaneously populating this one excited state. The dynamics that is triggered when the boost is applied to one molecule can therefore be seen as an approximation to the situation that the molecule is initially transferred into an excited state by absorption of a photon. We also stress that the computational approach that we present here is not tied to the boost excitation, and other, explicitly time-dependent excitations can 
be used as well.

Because we want to measure excitation-energy transfer from m5 to m1 we design an operator of the form (\ref{KS_bath_op}) such that it couples to variations of the dipole moment $|\mathbf{d}_1(t) - \mathbf{d}_1(t_0)|$ of molecule m1 only. The index indicates that the dipole moment is calculated only in the (45-degree) section of the grid corresponding to m1. 
The bath operator projects onto the ground state and thus removes the entire excitation energy.
This model is to be understood as an averaging approximation to a more realistic de-excitation mechanism in which the energy would dissipate to many lower energy levels. The scaling factor $M$ in the bath operator includes a normalization factor $D$ and reads
\begin{equation}
 M[n(\bfr,t)] = \frac{|\mathbf{d}_1(t)-\mathbf{d}_1(t_0)|}{D}.
 \label{bath_op_final}
\end{equation}
So far the time constant related to the rate of the dissipative mechanism is a free parameter, whereas $D$ needs to be chosen reasonably as discussed below in  \ref{bath_op_assessment_section}.
In summary, the bath mechanism measures locally the amount of excitation energy that has reached m1 and 
de-excites the entire system accordingly. 
The motivation for this concept will be further explained below.

In this first study our focus is on establishing our approach and on a transparent analysis of the influence of the intermolecular coupling mechanism. Therefore, although our setup is suitable for general chromophores in a ring-like arrangement, we here choose dimers as the components of the circular system (see \ref{model_system_appendix}). For these the electronic structure, e.g., excited-state energies, can easily be modified by bond-length variation. In this way, defects can be introduced in a well controlled way. For a transparent analysis we modify only molecules m3 and m7 in our investigations in  \ref{results_section} and fix all other system components, see  \ref{supermol_arrangement}. The electronic-structure properties of the model system are discussed in detail in \ref{model_system_appendix}.

In a circular arrangement of model molecules there are at least three sources for relevant time scales: time scales related to the energy of the individual molecular excited states, time scales due to the intermolecular coupling (here, more than one time scale is involved if the system is partly or totally off-resonant), and time scales due to the dissipative bath action. 
The first two types of time scales are determined by the molecular setup and system-internal interactions. The related dynamics are accessible in closed quantum-system simulations. Yet, for computing the energy-transfer dynamics of a system interacting with its environment, i.e. also taking the latter type of time scales into account, open quantum-system schemes as the one presented here are needed. 
We emphasize that Markovian system-bath couplings, as e.g. the one employed in the present approach, will only provide a damping mechanism which prevents
excitation energy from coherently travelling throughout the system. Besides a small Lamb-Shift type correction induced by the Markovian system-bath coupling, 
no modifications of excitation-energy transfer times can be achieved in such a coupling limit. Non-Markovian system-bath couplings on the other hand will be 
capable of modifying excitation-energy transfer times. As shown in Ref. \cite{gaspard:5676}, stochastic Schr\"odinger equations naturally allow to include 
Non-Markovian system-bath interactions. A natural next step beyond the present exploratory work is therefore to investigate Non-Markovian system-bath 
couplings within a stochastic TDKS approach.

Before we discuss the life-time of the excitation on the ring in \ref{results_section}, we first study the influence that the bath has on a single dimer.

\subsection{Assessment of the bath operator}
\label{bath_op_assessment_section}

The aim of this section is to define $D$ in such a way that the role of the scaling factors $\gamma$ and $(|\mathbf{d}_1(t)-\mathbf{d}_1(t_0)|)/D$ that appear in front of the projector can be well separated: $\gamma$ should set the decay time $\tau=1/\gamma$, and the dipole dependent scaling factor should account for the coupling to the dipole moment of molecule m1 without interfering with the role of $\gamma$. Therefore, $D$ needs to be adapted to the dipole oscillations of the isolated model molecule m1.

\begin{figure}
  \includegraphics[width=8.5cm]{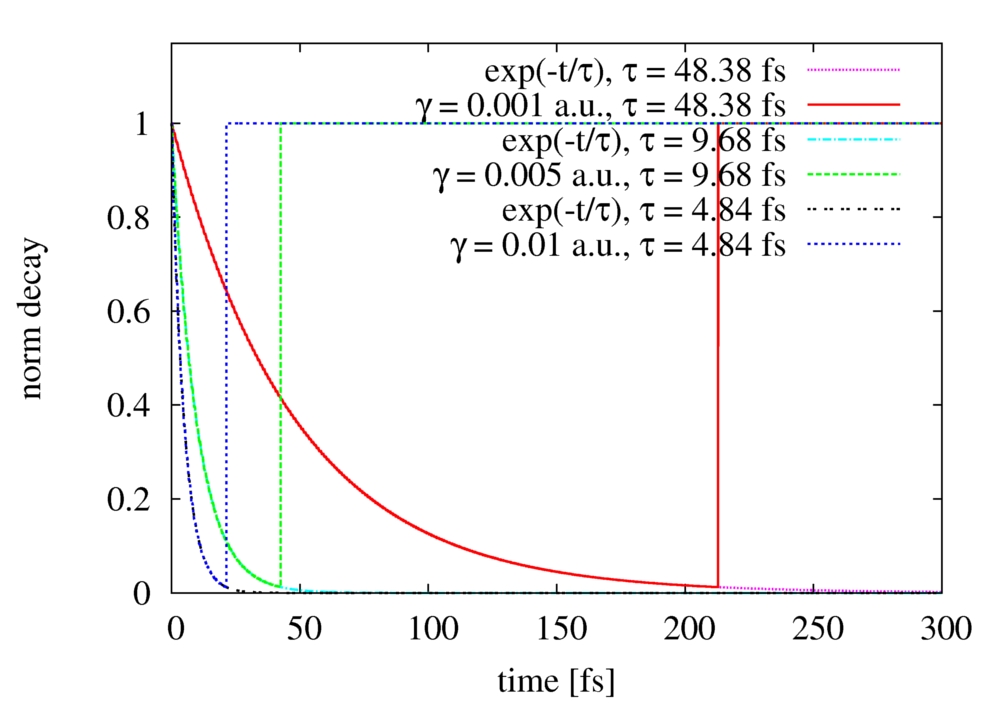}\\
  \caption{Norm decay $\eta(t)$ of a single model molecule. The damping is performed with different decay-time constants $\tau$ after an initial boost excitation with 0.001 eV excitation energy. The norm decay is always exponential $\exp(-t/\tau)$ with the preset time $\tau$. Here, quantum jumps were performed as the norm dropped below 0.014\ \%. They are seen as vertical lines leading back to a norm of one.}
\label{bath_op_damp_comp}
\end{figure}

To determine $D$ we calculated 100 fs of the dipole-moment time evolution of a single molecule as a closed quantum system after an initial momentum boost along the bond axis of the dimer. Thus, only the dipole moment along this axis is excited and the investigation can be restricted to this dipole-moment component. {\it A priori} one can think of different options for computing $D$: It may be chosen to be, e.g.,~the first maximum, the average over all maxima, the absolute average, or the absolute square average of the dipole moment. We found by numerical tests that an exponential decay with the time constant $\tau$ is reached only when $D$ is chosen as the average of the absolute square of the dipole moment.  \ref{bath_op_damp_comp} shows the norm decay of the auxiliary system of the quantum-jump algorithm. For all cases studied, the actually observed decay is exponential with the preset decay rate, as desired.

Another factor that needs to be taken into account in setting up $D$ is the energy of the boost excitation. The dissipation of energy should be independent of the boost strength. Therefore, we adapt the normalization factor to the boost strength by first determining $D$ from a closed quantum-system calculation with the same initial boost excitation that we later use in the open quantum-system calculation. Our calculations show that the normalization factor thus obtained yields a decay time that is independent of the boost strength. 

\begin{figure}
  \includegraphics[width=8.5cm]{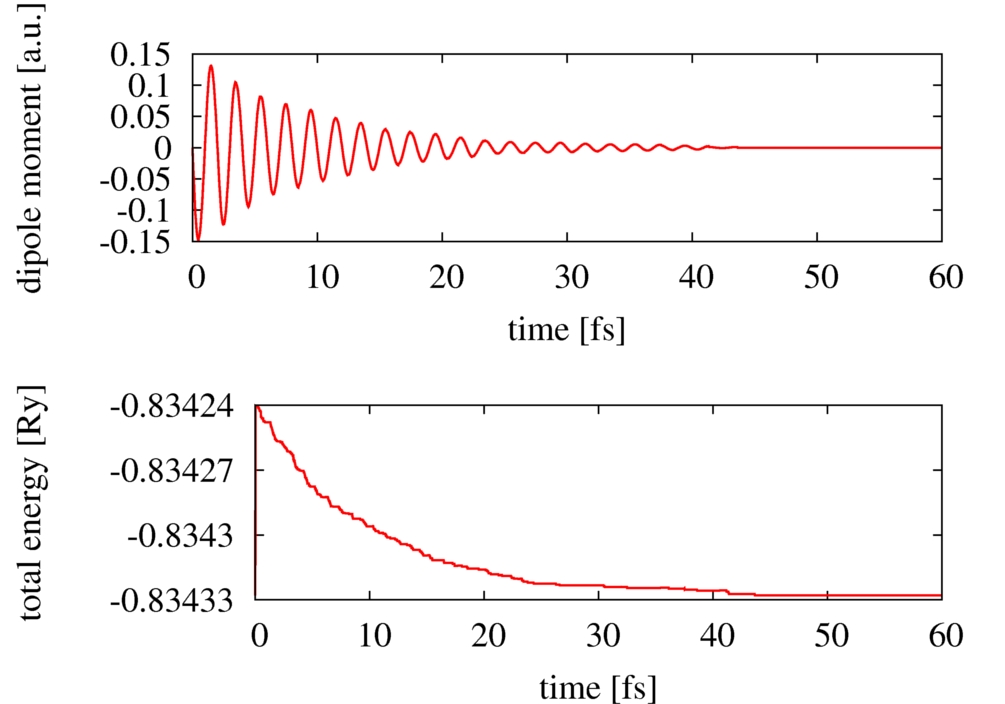}\\
  \caption{Ensemble-averaged dipole moment ($z$-component) and total energy of a single molecule in contact with a dissipative bath that induces a decay time of 10\ fs. 
  200 realizations have been computed.}
\label{bath_op_dipo_en}
\end{figure}

Having investigated the norm decay in detail, we finally give some results on other important observables. The exponential decay of the norm translates into an exponential decay of the total energy and the envelope of the dipole-moment oscillation (see  \ref{bath_op_dipo_en}). Thus, the bath operator fulfills all desired criteria.

\section{Excitation-energy spread in a ring system}
\label{results_section}

\subsection{Resonant excitation spread and decay time constants}
\label{interference_decay_time_const}

\begin{figure}
  \includegraphics[width=8.5cm]{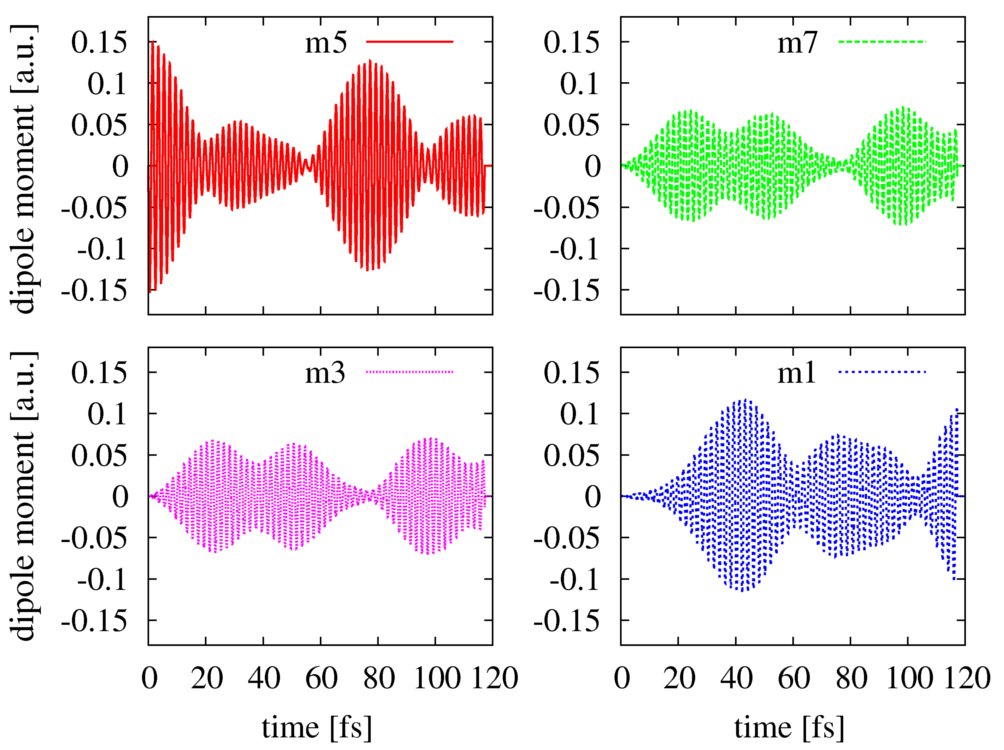}\\
  \caption{Dipole moment ($z$-component) of the four molecules m1, m3, m5, m7 in the ring system of  \ref{supermol_arrangement}. The intermolecular distance is 20\ Bohr. At $t=0$ m5 was excited by a 
  boost with 0.001 eV excitation energy.
  In this simulation there is no dissipation. An oscillation pattern emerges due to interferences of the dipolar excitation that is traveling around the ring.}
\label{resonant_dipole_evolution}
\end{figure}

We now investigate EET in circular supermolecules and start by studying the perfectly resonant coupling situation, i.e., all molecules of the ring are identical.  \ref{resonant_dipole_evolution} shows the time evolution of the dipole moment of four molecules in the ring of  \ref{supermol_arrangement}. The dynamics here is fully coherent, i.e., there is no coupling to the bath. The initial excitation is a boost of m5 as explained in  \ref{model_system_section}, and the intermolecular distance within the ring is 20\ Bohr. We observe a fast oscillation of the dipole moment that corresponds to the lowest excitation energy of our model system at 2.1\ eV, with an envelope that shows an interference pattern: At different points in time the largest dipole moment amplitude is observed on different subsystems. This interference pattern emerges as the dipolar excitation travels along the ring in both directions. It is governed by the intermolecular coupling strength, i.e., the pattern is determined by 
the time an excitation needs to be transferred between neighboring molecules. For a separation of 20\ Bohr the coupling strength is 0.038\ eV in our model system \cite{hofmann:012509}. If two of our model molecules were isolated, this coupling would amount to a resonance oscillation with a cycle duration of 107.8\ fs and after each quarter of this period the maximum of the dipole oscillation would be observed on one of the two neighbors \cite{hofmann:012509}. In the circular setup each molecule has neighbors on both sides and the excitation thus spreads through the ring.

\begin{figure}
  \includegraphics[width=8.5cm]{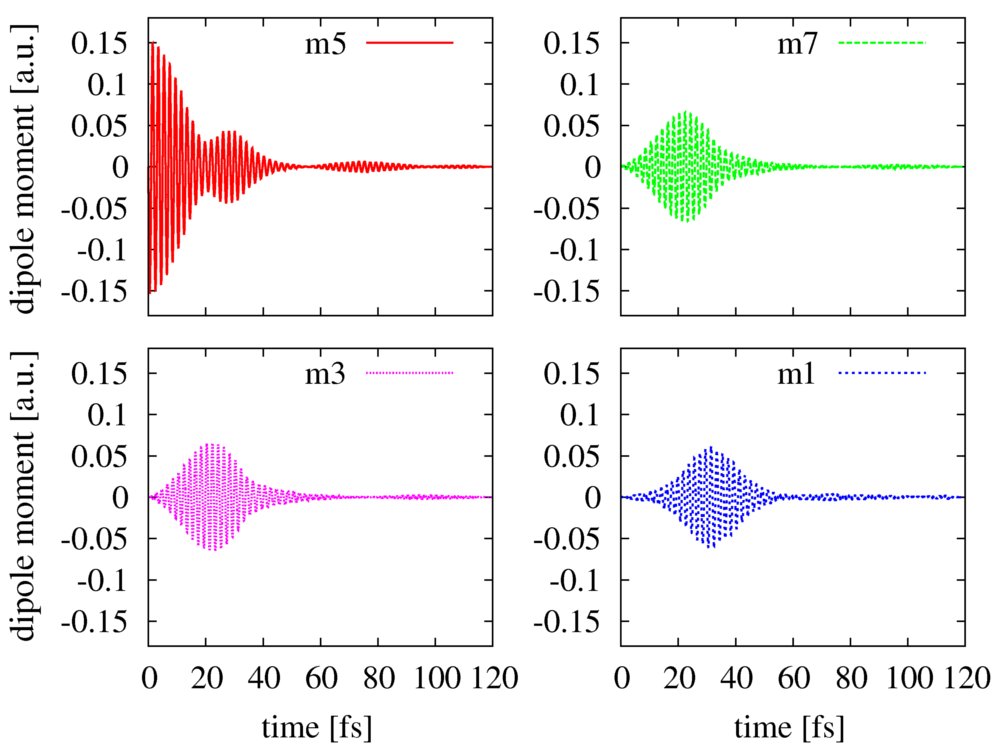}\\
  \caption{Ensemble-averaged (over 
  200 realizations) dipole moment ($z$-component) time evolution of four molecules that are arranged as in  \ref{resonant_dipole_evolution}. After the initial excitation at m5 the excitation travels in the ring and decays due to a dissipative bath that is acting with a decay time of 5\ fs on molecule m1.}
\label{damped_dipole_evolution}
\end{figure}

When we now add the bath mechanism to measure the EET time the bath needs to break the coherence and thus needs to operate on a comparably short time scale. We choose a decay time of 5\ fs, and this choice will be motivated further below.  \ref{damped_dipole_evolution} shows the time evolution of the dipole moment of the same molecules as in  \ref{resonant_dipole_evolution}. Due to the bath an additional time scale comes into play and one can clearly see how the oscillation of the dipole moment decays in all molecules due to the dissipative process. To be able to measure the traveling time of an excitation in the circular setup, the decay time needs to be chosen short enough to sample the initial stage of the coherent EET before the interference pattern starts to build up. In the example of  \ref{damped_dipole_evolution}, the decay time of 5\ fs fulfills this criterion: The dipole oscillation at molecule m1 reaches only one maximum and subsequently decays to the ground state. Thus, no 
interference emerges.

We define the time that serves as our measure for how long an excitation needs for traveling from m5 to m1 via the decay of the norm $\eta$ of the auxiliary system.
The rationale behind this choice is the following: The bath operator is sensitive to the (transition) dipole moment of m1, i.e., energy departs the system relative to the amount of excitation energy that arrives at molecule m1. The energy loss of the system is measured by the decay of the norm of the auxiliary system. Therefore, $\eta$ serves as measure for the excitation energy that reaches m1.
In practice we choose a threshold value for $\eta$ and monitor at which time $T$ the norm $\eta$ falls below this value. Depending on how large the chosen threshold is, the criterion can measure somewhat different effects. When the threshold is chosen relatively large (note that with a rapid norm decay as seen in \ref{bath_op_damp_comp} a value of 0.1 can be ``large''), then the time $T$ is determined by just the initial energy transfer and reflects how quickly energy spreads through a previously unexcited system. When the threshold is chosen smaller, than the criterion not only measures the initial energy transfer but also takes into account that the energy transfer can change as molecules on the way between m5 and m1 become excited. This way of defining energy-transfer times is not unique, but it appears reasonable on physical grounds. In our 
study we considered the norm decay down to three different thresholds, 0.100, 0.050, and 0.012. The first line of \ref{en_disorder_decay_times} lists the times $T$ that we determined for the ring with all molecules identical. These values serve as reference values for investigating how the excitation spread-time changes when defects are introduced into the ring.

\subsection{Influence of energetic disorder}
\label{influence_en_disorder}

\begin{table}
\begin{tabular}{lllll}
\hline
\hline\\[-3mm]
 m3 & m7 & $\eta=0.100$ & $\eta=0.050$ & $\eta=0.012$ \\
\hline
\ 0.0 Bohr & \ 0.0 Bohr  & \ \,52 fs & \ \,75 fs & 117 fs \\
 -0.1 Bohr & \ 0.0 Bohr  & \ \,56 fs &    110 fs & 121 fs \\
 -0.5 Bohr & \ 0.0 Bohr  & \ \,84 fs & \ \,97 fs & 143 fs \\
 -0.1 Bohr &  -0.1 Bohr  & \ \,57 fs &    112 fs & 123 fs \\
 -0.1 Bohr &  -0.5 Bohr  & \ \,85 fs & \ \,97 fs & 157 fs \\
 -0.5 Bohr &  -0.5 Bohr  &    166 fs &    181 fs & 211 fs \\
 removed   & \ 0.0 Bohr  & \ \,80 fs & \ \,89 fs & 125 fs \\
 removed   &  -0.1 Bohr  & \ \,99 fs &    125 fs & 147 fs \\
 removed   &  -0.5 Bohr  &    205 fs &    227 fs & 265 fs \\
\hline
\hline
\end{tabular}
\caption{\label{en_disorder_decay_times} Times $T$ where the norm $\eta(T)$ of the decaying auxiliary system drops below 0.100, 0.050, and 0.012. The different lines correspond to different setups of the ring. We modified the ring by either varying the bond length and thus the excitation energies of m3 and m7, or by removing them completely, as indicated in the first two columns of the table. A bond-length reduction of 0.1\ Bohr (0.5\ Bohr) amounts to an energetic detuning of the excitation energy by 0.049\ eV (0.135\ eV), and a reduction of the coupling-matrix element by 0.006\ eV (0.012\ eV), respectively.}
\end{table}

We assess the influence of energetic disorder on the EET time by introducing defects via bond-length variations and/or removing one of the molecules. The lower lines of  \ref{en_disorder_decay_times} list the times $T$ for the cases that a small (line 2) and a stronger energetic detuning (line 3) have been introduced in one arm of the ring, in both arms of the ring (lines 4 and 5), and with one molecule removed in one arm (lines 6 to 9). We find that small changes of the bond length (-0.1\ Bohr) in m3 or in both m3 and m7 result in only a moderate increase of the transfer time. Larger increases are observed when a larger defect (-0.5\ Bohr) is introduced in one of the molecules. The table also shows that interference effects have an influence on $T$: While the times corresponding to a decay down to $\eta=0.100$ and $\eta=0.012$ give a consistent picture, the situation is different for the times $T$ corresponding to $\eta=0.050$. Independent of this, however, there is a clear rise of the transfer time 
as soon as the more severe defect is introduced in both arms of the ring. In this case the life-time of the excitation on the ring increases by about a factor of two.

Another type of defect can be introduced by removing a molecule (m3) from the ring. In this case the coupling between molecules m2 and m4 is about an order of magnitude smaller than the next-neighbor coupling between the other molecules \cite{hofmann:012509}.
The first interesting observation in this case is made by comparing line 7 of  \ref{en_disorder_decay_times} (m3 removed) to line 3 (m3 disturbed): Detuning the excitation energy of m3 inhibits energy transfer more than removing m3 completely. The second observation is that energy transfer is seriously hindered when now also the other arm of the ring is disturbed by detuning molecule m7.

We summarize the findings by concluding that (i) interference effects play a role for the life-time of the excitation on the ring, (ii) as soon as sizable defects occur in both pathways EET is hindered noticeably, whereas a defect in one pathway has less dramatic influences, (iii) for resonant or close-to-resonant cases the EET time is not affected much by variations of the coupling strength, and removing one subsystem of the ring hinders EET less than a sizeable energetic detuning of this subsystem.

\subsection{Influence of the intra-system coupling}
\label{influence_intrasys_coupling}

In the final investigation we examine how EET times change when the frequently employed \cite{scholes:57, konig:386, hofmann:012509} dipole-coupling approximation of F\"orster theory is used instead of the full coupling. For our model system it is known\cite{hofmann:012509} that the true coupling between two molecules differs noticeably from the dipole-dipole approximation for distances below 20\ Bohr. Therefore, the intermolecular distance was reduced to 12\ Bohr in the following calculations to model a situation in which the ring members are so close that the dipole approximation cannot be expected to be realistic. 
We implement the dipole-coupling approximation by a multipole expansion of the molecular interaction and truncating the latter after the dipole-dipole contribution. More details are described in Ref.\ \cite{hofmann:012509}. 
Here, we use F\"orster-type dipole-coupling between all molecules in the ring setup.
The aim of this study is to check which conclusions one would (wrongly) draw when one assumes dipole coupling in a situation where the dipole approximation breaks down. The (full) coupling-matrix element at 12\ Bohr is 0.105\ eV \cite{hofmann:012509}, i.e., notably larger than in the previous examples, but we kept the dissipation time constant at 5\ fs.

\begin{table}
\begin{tabular}{lllll}
\hline
\hline\\[-3mm]
 coupling & m3 & $\eta=0.100$ & $\eta=0.050$ & $\eta=0.012$ \\
\hline
 full       & \ 0.0 Bohr & \ \,95 fs &    116 fs & 163 fs \\
 F\"orster  & \ 0.0 Bohr & \ \,86 fs &    104 fs & 152 fs \\
 full       &  -0.1 Bohr & \ \,99 fs &    140 fs & 182 fs \\
 F\"orster  &  -0.1 Bohr & \ \,94 fs &    108 fs & 154 fs \\
 full       &  -0.5 Bohr &    109 fs &    140 fs & 209 fs \\
 F\"orster  &  -0.5 Bohr & \ \,89 fs &    115 fs & 185 fs \\
\hline
\end{tabular}
\caption{\label{intrasys_coupling_decay_times} Compiled are the times $T$ at which $\eta(T)$ drops below 0.100, 0.050, and 0.012 (columns 3, 4, and 5, respectively) for a ring with an intermolecular distance of 12\ Bohr. Full coupling is compared to dipole-dipole coupling. The ring is disturbed via bond-length variations in molecule m3 as indicated in the second column.}
\end{table}

 \ref{intrasys_coupling_decay_times} shows the EET times that we determined as described above for an undisturbed ring and with perturbations on one side of the ring. In each case we ran a simulation once with full coupling and once with dipole-dipole coupling but otherwise identical parameters. The table shows that the F\"orster-type dipole coupling underestimates the times $T$ in all cases, i.e., in our setup it overestimates EET efficiency. This finding is in line with the observation that the coupling-matrix element is overestimated by the dipole approximation \cite{hofmann:012509}. A further interesting observation is that the times $T$ here, i.e., for an intermolecular distance of 12\ Bohr, are larger than the corresponding times for an intermolecular distance of 20\ Bohr, although in the latter case the coupling is notably weaker. This may seem like a contradiction, but is a real effect caused by the interferences that are more pronounced in case of stronger coupling. Therefore, care has to be taken 
when comparing 
absolute times unless the dissipation time constant of the bath operator is chosen short enough to prevent a buildup of interferences due to excitation-energy spread along different available pathways.

\section{Summary and Conclusions}
\label{summary_section}

In summary, we developed and used a stochastic TDDFT-based approach for open quantum systems to investigate the excitation-energy spread in a circular arrangement of molecules. We introduced a bath operator that couples to the dipole moment of specific subsystems of the supermolecular complex. The dissipative mechanism breaks the coherent EET and removes excitation energy from the system. It allows for measuring energy-transfer times. The influence of electronic-structure properties and intermolecular coupling mechanisms on the EET process can thus be investigated.
We demonstrated among other things that sizeable perturbations of the electronic structure of a model ring system lead to a considerable decrease in energy-transfer efficiency if both pathways of the ring are perturbed, completely removing a molecule from the ring can inhibit energy transfer less than having an energetically detuned molecule in the ring, and F\"orster's dipole coupling approximation 
may overestimate EET efficiency noticeably. Our scheme is completely general and future applications can be extended to much more complex molecular setups than the one of this first study. The approach can therefore help to shed light onto the complex phenomena that govern one of nature's most fascinating processes, the collection of light by plants and bacteria.

Financial support by the DFG Graduiertenkolleg 1640 is gratefully acknowledged by SK and DH, and DH also acknowledges the hospitality of the UCSD. MD acknowledges support from DOE grant DE-FG02-05ER46204.

\begin{appendix}

\section{Single-particle representation of the bath operator}
\label{sp_bath_op_appendix}

In this appendix we demonstrate that using the single-particle operator $\hat{s}_i$ of (\ref{KS_bath_op}) in the set of single-particle equations of \ref{simu_algo} is equivalent to solving (\ref{stKSEq}) with the bath operator $\hat{S}$ of  (\ref{bath_op_scaled}) using the quantum-jump algorithm. In the latter case, one needs to propagate simultaneously the norm-preserving equation
\begin{equation}
 \mathrm i \partial_t \Phi_\mathrm{KS} = \hat{H}_\mathrm{KS}\Phi_\mathrm{KS}-\frac{\mathrm i}{2} \hat{S}^\dagger \hat{S} \Phi_\mathrm{KS} + \frac{\mathrm i}{2} ||\hat{S}\Phi_\mathrm{KS}||^2 \Phi_\mathrm{KS}
 \label{qj_eq}
\end{equation}
and the auxiliary equation
\begin{equation}
 \mathrm i \partial_t \Phi_\mathrm{KS}^\mathrm{aux} = \hat{H}_\mathrm{KS}\Phi_\mathrm{KS}^\mathrm{aux}-\frac{\mathrm i}{2} \hat{S}^\dagger \hat{S} \Phi_\mathrm{KS}^\mathrm{aux}.
 \label{qj_eq_decay}
\end{equation}
Inserting $\hat{S}$ of  (\ref{bath_op_scaled}) in  (\ref{qj_eq}) yields the closed quantum-system KS equation
\begin{equation}
 \mathrm i \partial_t \Phi_\mathrm{KS} = \hat{H}_\mathrm{KS}\Phi_\mathrm{KS}.
\end{equation}
The latter equation can be translated to the well-known set of single-particle KS equations that one also obtains by inserting $\hat{s}_i$ of  (\ref{KS_bath_op}) into  (\ref{stSEq_norm_cons}).

It remains to be checked that with our choice of bath operator the single-particle equation (\ref{stSEq_norm_decay}) indeed corresponds to the many-particle equation (\ref{qj_eq_decay}). To this end we use the method of Ref.\ \cite{wong:1419} for deriving the single-particle equations that correspond to  (\ref{qj_eq_decay}). We start from the reduced one-body density matrix
\begin{equation}
\begin{split}
 \rho(\bfr,\bfr',t) = \int & \Phi_\mathrm{KS}(\bfr,\bfr_2,\ldots,\bfr_N) \\
 & \Phi^*_\mathrm{KS}(\bfr',\bfr_2,\ldots,\bfr_N)\,d^3 r_2 \ldots d^3 r_N
\end{split}
\end{equation}
and compute the equation of motion for it by inserting the open quantum-system KS equation and the specific form of  (\ref{bath_op_scaled}) for $\hat{S}$  to find
\begin{equation}
\begin{split}
 \mathrm i & \partial_t \rho(\bfr,\bfr',t) = \\ & \left[ \hat{h}_\mathrm{KS}(\bfr,t)-\frac{\mathrm i}{2}\gamma M^2- \left( \hat{h}_\mathrm{KS}(\bfr',t)+\frac{\mathrm i}{2}\gamma M^2\right) \right] \rho(\bfr,\bfr',t),
 \label{eq_motion_rho}
\end{split}
\end{equation}
where $M$ follows from  (\ref{bath_op_scaled}).
Next, we express $\rho(\bfr,\bfr',t)$ in terms of a representation of $N$ orthonormal basis functions $\tilde{\varphi}_j$ \cite{wong:1419},
\begin{equation}
 \rho(\bfr,\bfr',t) = \sum_{j=1}^N \tilde{\varphi}_j(\bfr,t) \tilde{\varphi}_j^*(\bfr',t).
\end{equation}
Inserting this representation into  (\ref{eq_motion_rho}) yields
\begin{equation}
\begin{split}
 \sum_{j=1}^N \tilde{\varphi}_j^*(\bfr',t) \left[ \mathrm i \partial_t-\hat{h}_\mathrm{KS}(\bfr,t) \right] \tilde{\varphi}_j(\bfr,t) = \\
 \sum_{j=1}^N \left\{ \tilde{\varphi}_j^*(\bfr,t) \left[ \mathrm i \partial_t-\hat{h}_\mathrm{KS}(\bfr',t) \right] \tilde{\varphi}_j(\bfr',t)\right\}^*
 \label{hermiticity}
\end{split}
\end{equation}
for the case without dissipation ($\gamma = 0$).
In the limit $\gamma = 0$ a single-particle representation can be obtained by introducing the time-dependent matrix
\begin{equation}
 C_{jj'}(t) = \left\langle \tilde{\varphi}_j(t) \left| \mathrm i \partial_t- \hat{h}_\mathrm{KS}    (\bfr,t) \right| \tilde{\varphi}_{j'}(t) \right\rangle.
\end{equation}
Equation (\ref{hermiticity}) demonstrates that the matrix $C_{jj'}(t)$ is Hermitian (see also Ref.\ \cite{wong:1419}) and, therefore, can be diagonalized. Based on this finding, we assume that $\{ \tilde{\varphi}_j(\bfr,t) \}$ is the diagonal basis, without loss of generality, and that it fulfills the eigenvalue equation
\begin{equation}
 \left[ \mathrm i \partial_t-\hat{h}_\mathrm{KS}    (\bfr,t) \right] \tilde{\varphi}_j(t) = \varepsilon_j(t)\tilde{\varphi}_j(t)
\end{equation}
with real eigenvalues $\varepsilon_j(t)$. Next, we consider the case which includes dissipation. For $\gamma \ne 0$ we can introduce in a similar way a matrix $D_{jj'}(t)$ by
\begin{align}
 D_{jj'}(t) & = \left\langle \tilde{\varphi}_j(t) \left| \mathrm i \partial_t- \left(\hat{h}_\mathrm{KS}(\bfr,t) -\frac{\mathrm i}{2}\gamma M^2  \right)\right| \tilde{\varphi}_{j'}(t) \right\rangle \nonumber \\
            & = C_{jj'}(t) + \frac{\mathrm i}{2}\gamma M^2 \left\langle \tilde{\varphi}_j(t) | \tilde{\varphi}_{j'}(t) \right\rangle,
\end{align}
which includes the dissipative term. $D_{jj'}(t)$ contains the Hermitian contribution $C_{jj'}(t)$ and
an anti-Hermitian part $\frac{\mathrm i}{2}\gamma M^2 \left\langle \tilde{\varphi}_j(t) | \tilde{\varphi}_{j'}(t) \right\rangle$.
Since our dissipation operator is proportional to a unit operator, the matrix $D_{jj'}(t)$
also remains diagonal in the basis $\{ \tilde{\varphi}_j(\bfr,t) \}$. The eigenvalues $\epsilon_j(t)$ of $D_{jj'}(t)$ differ from the ones of $C_{jj'}(t)$ by a constant shift in the imaginary part,
\begin{equation}
 \epsilon_j(t) = \varepsilon_j(t) + \frac{\mathrm i}{2}\gamma M^2.
\end{equation}
We can set
\begin{equation}
 \tilde{\varphi}_j(t) = \exp\left\{ -\mathrm i \int_{t_0}^{t} \epsilon_j(\tau)\,d\tau \right\} \varphi^\mathrm{aux}_j(t)
\end{equation}
to obtain the set of single-particle equations
\begin{equation}
\label{single_particle_plus_dissipation}
 \mathrm i \partial_t \varphi_j^\mathrm{aux}(t) = \hat{h}_\mathrm{KS}\varphi_j^\mathrm{aux}(t)-\frac{\mathrm i}{2} \gamma M^2 \varphi_j^\mathrm{aux}(t).
\end{equation}
Note that due to the dissipation, the orbitals $\varphi_j^\mathrm{aux}(t)$ do not stay normalized during the time-evolution.\\
In summary, we have thus demonstrated that the single particle equations in (\ref{single_particle_plus_dissipation}) are equivalent to the set of equations\ (\ref{stSEq_norm_decay}) when the operator $\hat{s}_i$ is inserted.

\section{Details of the model system}
\label{model_system_appendix}

To guarantee a transparent analysis of the coupling and energetic arrangement we chose sodium dimers as well-accepted model molecules \cite{hofmann:012509,calc_info}. The Na dimer exhibits strong dipolar character and the electronic structure can easily be modified by bond-length variations (the experimental bond length is 5.78 Bohr). All sodium dimers are aligned along the $z$-axis according to the setup of  \ref{supermol_arrangement} and their centers of mass are in the $x$-$y$-plane. In the following we discuss the electronic-structure and coupling properties of our model system.

\subsection{Single molecule electronic structure}
\label{subsystem_electronic_structure_section}

\begin{figure}
  \includegraphics[width=8.5cm]{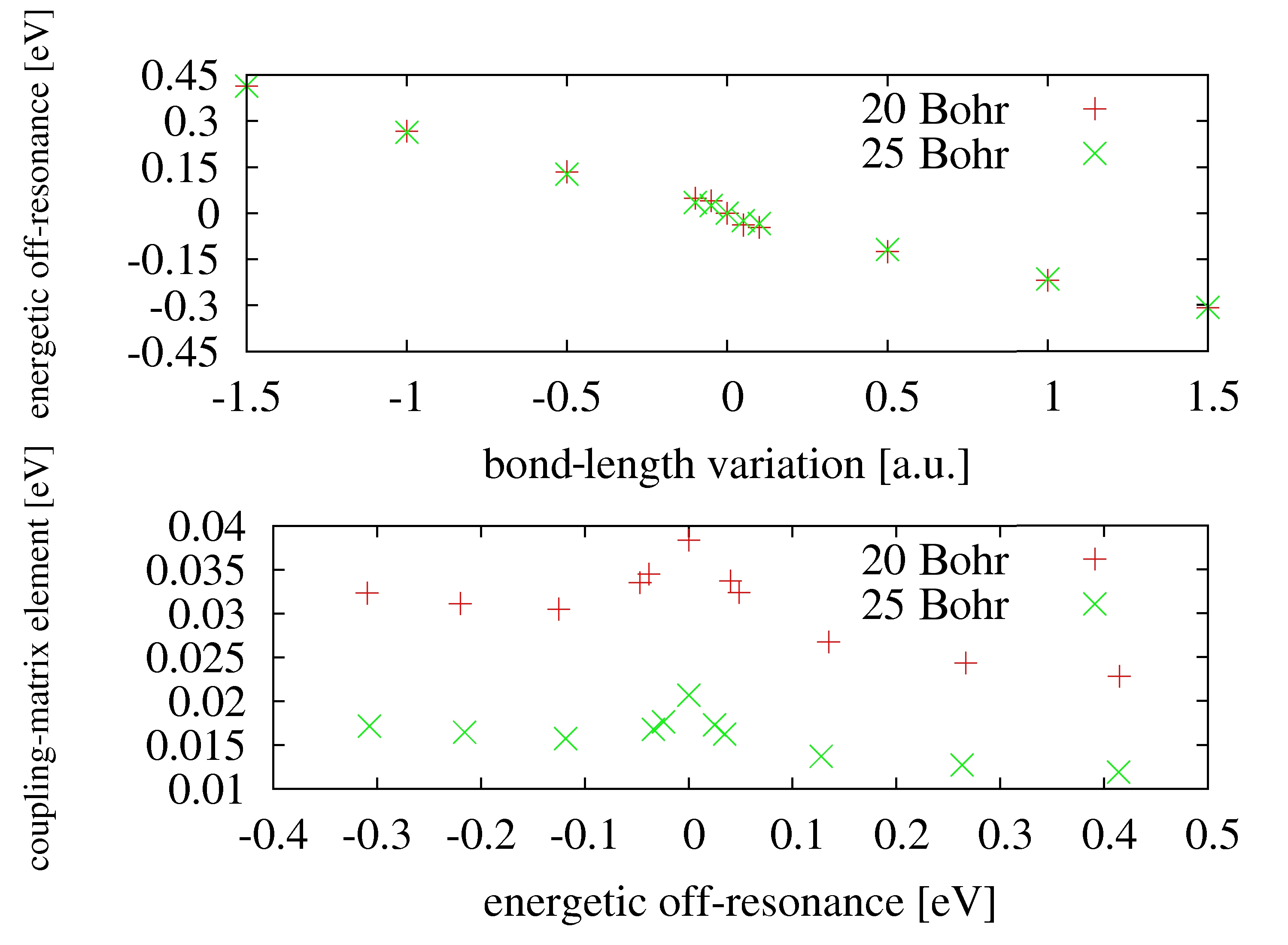}\\
  \caption{Upper panel: 
  Excitation energy
  (energetic off-resonance) of Na$_2$ as a function of the bond length as computed in the supersystem calculation explained in \ref{coupling_sec}. Lower panel: Coupling-matrix element calculated in a supersystem of two Na$_2$ as a function of the energetic off-resonance of one of the dimers.}
\label{el_struc_coupling_fig}
\end{figure}

First, we discuss the electronic structure of one dimer, i.e., one subsystem in the ring. We start with Na$_2$ at the experimental bond length. For excitations oriented along the bond axis, Na$_2$ is almost a single level atom as there is one prominent excitation at ca.\ 2.1 eV
and a second excitation at 4.1 eV with notably smaller oscillator strength.
The excitation energy of the one prominent peak can be calculated as a function of the bond length in two different ways: either from straightforward TDDFT for a dimer with changed bond length, or in a supersystem setup with two dimers (see  \ref{coupling_sec}). We used both methods and found reasonable agreement. The upper panel of \ref{el_struc_coupling_fig} shows that bond length changes of the order of 1 Bohr induce excitation-energy changes of about 0.25 eV.

\subsection{Resonant and off-resonant coupling}
\label{intermol_coupling_section}

Second, we study the coupling between two neighboring molecules. Here, we address two influencing factors, namely the distance between the dimers and the energetic arrangement determined by the dimer bond length. The distance dependence of the coupling-matrix element in a resonant situation with sodium dimers of equal bond length was already investigated in an earlier work \cite{hofmann:012509}. For distances above 25 Bohr, the coupling is of dipole-dipole type, whereas clear deviations from the dipole-dipole character can be observed for smaller distances.

In an off-resonant coupling situation between isolated excited states, a similar two-level picture \cite{neugebauer:2207} as in the resonant case  \cite{hofmann:012509} can be used to determine the coupling-matrix element. In analogy to the resonant case the coupling strength can be extracted from the dipole moment of separate system parts (for details, see \ref{coupling_sec}). The coupling-matrix element as a function of the energetic off-resonance of the excitation energy of one of the dimers is shown in the lower panel of  \ref{el_struc_coupling_fig}. In the range that we investigated, the coupling-matrix element decreases with increasing excitation energy. The coupling strength peaks at the resonant coupling situation, and the resonant coupling is about 1.4 times larger than the typical off-resonant coupling.

\section{Off-resonant oscillation in the time-dependent dipole moment}
\label{coupling_sec}

In this appendix we briefly explain how the coupling strength and the energetic off-resonance manifest in the time-dependent dipole moment for a system of one donor molecule D and one acceptor molecule A. The resonant coupling case was already discussed in Ref. \cite{hofmann:012509}. Our considerations employ a two-state model \cite{neugebauer:2207} based on the assumption that the wave function of the total system can be separated into D and A parts due to negligible electronic overlap between D and A. Initially, the acceptor is in its ground state denoted by $|\mathrm{A}\rangle$ and the donor is in an excited state $|\mathrm{D}^*\rangle$. This corresponds to an initial product state $|\mathrm{D}^*\mathrm{A}\rangle=|\mathrm{D}^*\rangle|\mathrm{A}\rangle=|1\rangle$. The final
wave function corresponds to the inverse situation, $|\mathrm{D}\mathrm{A}^*\rangle=|2\rangle$. Both states are characterized by their eigenvalues $E_1$ and $E_2$, respectively. We measure energetic off-resonances by the parameter 
\begin{equation}
\Delta E=\frac{1}{2}(E_1-E_2).
\end{equation}
The coupling between $|1\rangle$ and $|2\rangle$ is mediated by the Coulomb interaction $\hat{V}_\mathrm{C}$. It leads to the coupling-matrix element
\begin{equation}
V=\langle \mathrm{D}\mathrm{A}^*|\hat{V}_\mathrm{C}|\mathrm{D}^*\mathrm{A}\rangle.
\label{coupling}
\end{equation}

The time evolution of the two-state system with initial state $|\Psi(0)\rangle=|1\rangle$ is given by
\begin{equation}
|\Psi(t)\rangle=a_1(t)|1\rangle+a_2(t)|2\rangle
\end{equation}
with the coefficients
$a_{1}(t)$ and $a_{2}(t)$ \cite{cohen-tannoudji1:2009},
\begin{equation}
\begin{split}
|a_1(t)|^2 & = B + A \cos^2\left(\sqrt{V^2+\Delta E^2}t\right), \\
|a_2(t)|^2 & = A \sin^2\left(\sqrt{V^2+\Delta E^2}t\right),
\label{two_level_coeff}
\end{split}
\end{equation}
where $A = \frac{V^2}{V^2+\Delta E^2}$ and $B = \frac{\Delta E^2}{V^2+\Delta E^2}$. This time evolution of the coefficients corresponds to an incomplete oscillation with beat frequency $\omega_\mathrm{beat}= \sqrt{V^2+\Delta E^2}$ that depends on the coupling between the initial and the final state as well as the energetic off-resonance: The occupation probability of the initial state varies around B with amplitude A, while the occupation probability of the final state oscillates with amplitude A around zero.

\begin{figure}
  \includegraphics[width=8.5cm]{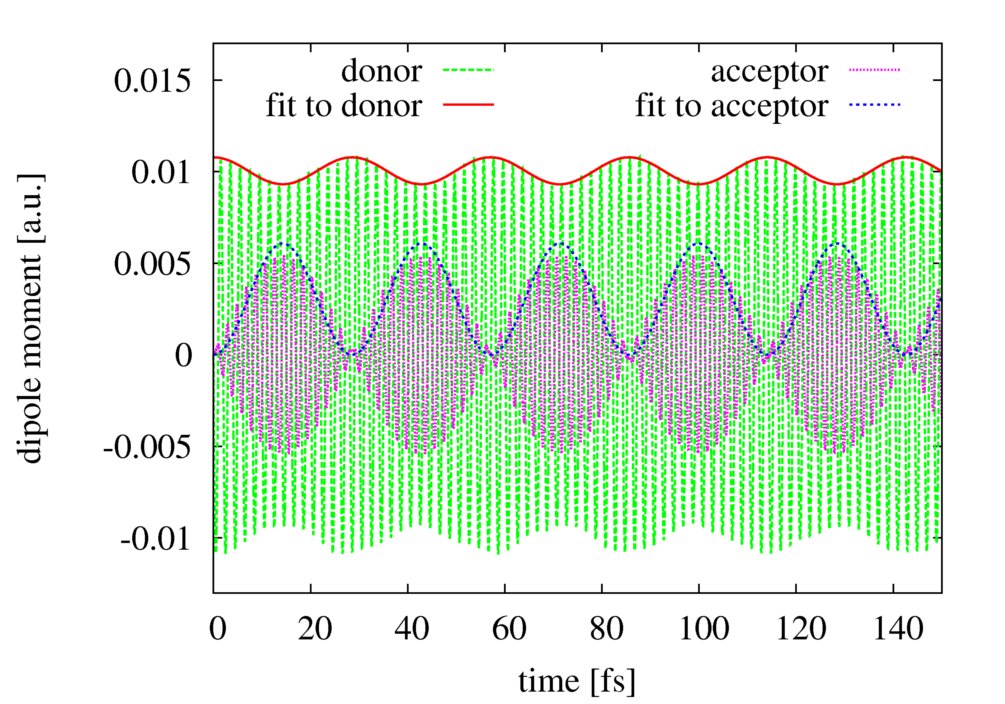}\\
  \caption{Donor and acceptor dipole moment ($z$-component) in a setup of two Na$_2$, where the bond length of the acceptor Na$_2$ is reduced by 0.5 Bohr compared to the experimental bond length. We performed fits to the envelope of the oscillation according to the model discussed in the text.}
\label{dipo_off-res_osc_fig}
\end{figure}

The time-dependent dipole moment $\textbf{d}^\mathrm{A}(t)=\langle\Psi(t)|\hat{\textbf{r}}^\mathrm{A}|\Psi(t)\rangle$ of the acceptor, where we evaluate the dipole operator $\hat{\textbf{r}}^\mathrm{A}$ in the space of the acceptor only, can be calculated as
\begin{equation}
\textbf{d}^\mathrm{A}(t)=|a_1(t)|^2 \langle \mathrm{A}|\hat{\textbf{r}}^\mathrm{A}|\mathrm{A}\rangle+|a_2(t)|^2 \langle \mathrm{A}^*|\hat{\textbf{r}}^\mathrm{A}|\mathrm{A}^*\rangle.
\label{A_dip_mom1}
\end{equation}
Here, we exploited the orthogonality of $|D\rangle$ and $|D^*\rangle$. If the static dipole moment $\langle A|\hat{\textbf{r}}^\mathrm{A}|A\rangle$ of
A vanishes, (\ref{A_dip_mom1}) simplifies to
\begin{equation}
\textbf{d}^\mathrm{A}(t)=|a_2(t)|^2 \langle \mathrm{A}^*|\hat{\textbf{r}}^\mathrm{A}|\mathrm{A}^*\rangle.
\label{A_dip_mom2}
\end{equation}
Under the assumption that the static donor dipole moment vanishes, the donor time-dependent dipole moment reads
\begin{equation}
\textbf{d}^\mathrm{D}(t)=|a_1(t)|^2 \langle \mathrm{D}^*|\hat{\textbf{r}}^\mathrm{D}|\mathrm{D}^*\rangle.
\label{D_dip_mom}
\end{equation}
Thus, the resonance oscillation of the coefficients can be observed in the time
evolution of the dipole moments $\textbf{d}^\mathrm{A}(t)$ and $\textbf{d}^\mathrm{D}(t)$. In principle, both (\ref{A_dip_mom2}) and (\ref{D_dip_mom}) can be used to determine the coupling-matrix element $V$. Of special importance is  (\ref{D_dip_mom}) together with  (\ref{two_level_coeff}), because a fit to the absolute values of the extrema of the $k$-th component of the donor dipole moment time evolution provides $A p_k$, $B p_k$, and $\omega_\mathrm{beat}$, where $p_k=\langle \mathrm{D}^*|\hat{r}_k^\mathrm{D}|\mathrm{D}^*\rangle$. It can be used to determine $V$, $\Delta E$, and $p_k$. One obtains
\begin{equation}
   V = \sqrt{\frac{Ap_k}{Ap_k+Bp_k}}\omega_\mathrm{beat}
\end{equation}
and
\begin{equation}
   \Delta E = \sqrt{\frac{Bp_k}{Ap_k+Bp_k}}\omega_\mathrm{beat}.
\end{equation}

A typical time evolution of the $z$-component of the acceptor and donor dipole moments of our model system is depicted in  \ref{dipo_off-res_osc_fig} together with our fits to the envelopes. The two-level model qualitatively fits to the observed dipole oscillation of the off-resonant, coupled system of two molecules. However, the dipole moment envelopes do not perfectly follow the $\sin^2$- and $\cos^2$-shape. We understand these deviations as a consequence of the coupled system not being perfectly separable into D and A parts, as was assumed in the model. Furthermore, although the second excitation of Na$_2$ with polarization in z-direction is energetically far off and carries notably smaller oscillator strength, Na$_2$ is not a perfect single level system, and therefore the system of two sodium dimers does not perfectly match the two-level model. Nevertheless, our approach provides a tool to determine the coupling $V$ and the energetic off-resonance $\Delta E$ with reasonable accuracy. The validity of 
the model can be checked by comparing the $\Delta E$ thus obtained to TDDFT excitation energies for a single sodium dimer with changed bond length. We find good agreement and, therefore, assume reasonable quality of the coupling-matrix element results of  \ref{el_struc_coupling_fig}.

\end{appendix}

\newpage

  \includegraphics[width=8.9cm]{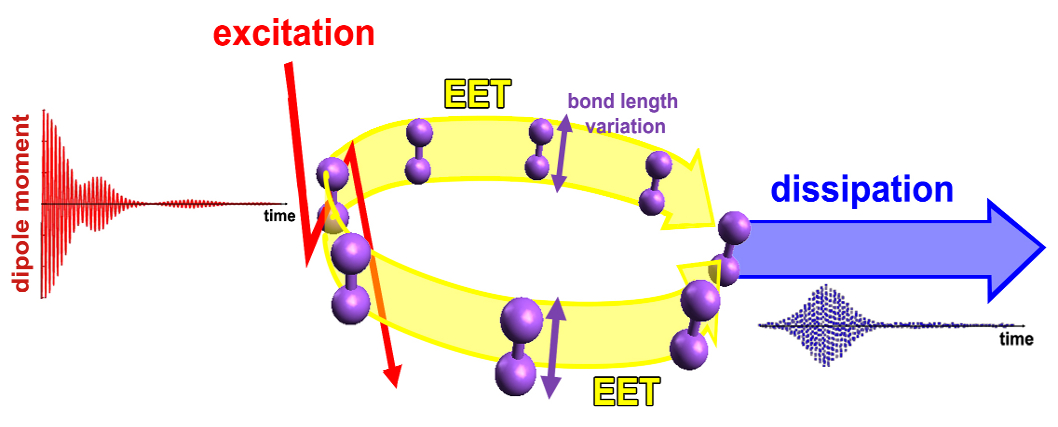}\\
Table of Contents figure: Illustration of a scheme to compute excitation energy transfer

\end{document}